\documentclass[aps,twocolumn,showkeys,amsmath,amssymb,superscriptaddress,showpacs]{revtex4}
\usepackage{graphicx} % Include figure files
\usepackage{color}
\usepackage{ulem} 
\newcommand{\lacorh}{LaCo$_{1-x}$Rh$_{x}$O$_3$}
\newcommand{\laco}{LaCoO$_3$}
\newcommand{\larh}{LaRhO$_3$}
\newcommand{\lnco}{$Ln$CoO$_3$}

\newcommand{\abbo}{$A_2BB'O_6$}
\newcommand{\lacotm}{LaCo$_{1-x}$M$_{x}$O$_3$}

\newcommand{\rtric}{$R\bar{3}c$}
\newcommand{\st}{$\rm ^o$}
\newcommand{\stc}{$\rm ^o$C}
\newcommand{\etal}{\textit{et al.}}

\newcommand{\muvk}{$\mu$V/K}

\begin{document}
\sloppy
\title{Stabilization of the high-spin state of Co$^{3+}$ in LaCo$_{1-x}$Rh$_{x}$O$_3$.}
\author{K. Kn\'{\i}\v{z}ek}
\email[corresponding author:]{knizek@fzu.cz}
\author{J. Hejtm\'{a}nek}
\author{M. Mary\v{s}ko}
\author{Z. Jir\'{a}k}
\affiliation{
 Institute of Physics ASCR, Cukrovarnick\'a 10, 162 00 Prague 6, Czech Republic.}
\author{J. Bur\v{s}\'{\i}k}
\affiliation{
 Institute of Inorganic Chemistry ASCR, 250 68 \v{R}e\v{z} near Prague, Czech Republic.}
\begin{abstract}
The rhodium doping in the LaCo$_{1-x}$Rh$_{x}$O$_3$ perovskite series ($x=0.02-0.5$) has been
studied by X-ray diffraction, electric transport and magnetization measurements, complemented by
electronic structure GGA+U calculations in supercell for different concentration regimes. No
charge transfer between Co$^{3+}$ and Rh$^{3+}$ is evidenced. The diamagnetic ground state of
LaCoO$_3$, based on Co$^{3+}$ in low-spin (LS) state, is disturbed even by a small doping of Rh.
The driving force is the elastic energy connected with incorporation of a large Rh$^{3+}$ cation
into the matrix of small LS Co$^{3+}$ cations, which is relaxed by formation of large Co$^{3+}$ in
high-spin (HS) state in the next-nearest sites to the inserted Rh atom. With increasing
temperature, the population of Co$^{3+}$ in HS state increases through thermal excitation, and a
saturated phase is obtained close to room temperature, consisting of a nearest-neighbor
correlation of small (LS Co$^{3+}$) and large (HS Co$^{3+}$ and LS Rh$^{3+}$) cations in a kind of
double perovskite structure. The stabilizing role of elastic and electronic energy contributions
is demonstrated in supercell calculations for dilute Rh concentration compared to other dopants
with various trivalent ionic radius.
\end{abstract}
%75.30.Wx    Spin crossover
%71.15.Mb    Density functional theory, local density approximation, gradient and other corrections
\maketitle
\pagestyle{myheadings} \markright{K. Kn\'{\i}\v{z}ek \etal,
 Stabilization of the high-spin state of Co$^{3+}$ in LaCo$_{1-x}$Rh$_{x}$O$_3$.}

\section{Introduction}

The cobalt perovskites \lnco\ ($Ln$~=~La, Y, rare-earths) show various anomalous behaviors that
are associated with changes of the spin state of octahedrally coordinated Co$^{3+}$ ions of $3d^6$
configuration. According to recent interpretation in the frame of LS-LS/HS-IS model, the process
can be understood as a local excitation of Co$^{3+}$ ions from the diamagnetic LS (low spin,
$t_{2g}^6$) ground state to closely lying paramagnetic HS (high spin, $t_{2g}^4 e_g^2$) states,
followed at higher temperature by a formation of a metallic-like phase of IS character
(intermediate spin, $t_{2g}^{5} \sigma^*$) - see \textit{e.g.}
\cite{RefKyomen2005PRB71_024418,RefJirak2008PRB78_014432,RefKnizek2009PRB79_014430}. In \laco\
these two steps are well separated ($T_{magn}=70$~K, $T_{I-M}=535$~K). The experiments show that
the HS population, after a steep increase at $T_{magn}$, tends quickly to a saturation, reaching
40-50\% in maximum - see \textit{e.g.} the combined analysis of susceptibility and anomalous
expansion data in \cite{RefKnizek2006JPCM18_3285}. To account for such behavior, two formal
approaches have been used. In the first one, Ky\^{o}men~\etal\ treat the excitations as locally
independent events obeying Boltzmann statistics and relate the actual course of the
diamagnetic-paramagnetic transition to lattice effects. An activation energy increasing
progressively with HS population is obtained \cite{RefKyomen2005PRB71_024418}. An alternative
approach is based on original idea of Goodenough that the stabilization of HS state is conditioned
by presence of LS states in the neighborhood. As shown by Kn\'{\i}\v{z}ek~\etal\
\cite{RefKnizek2006JPCM18_3285} using a simple probabilistic model, when nearest HS neighbors are
forbidden, the activation energy found by the fit has an opposite trend - it decreases with
temperature (or equivalently with HS population) leading finally to a crossover. Such behavior
finds a strong support in GGA+U calculations \cite{RefKnizek2009PRB79_014430} and also the
limiting HS population of about 45\% can be naturally explained.

An uncertainty exists also concerning the formation of metallic phase at around 535~K. The gradual
course of this process, which is in striking contrast with standard I-M transition, has been
modeled as a thermal excitation to IS states \cite{RefKnizek2008JAP103_07B703}. However, it is
worth mentioning that the two-step transitions in \laco\ does not necessarily require an existence
of three close lying ionic configurations (LS, HS and IS). A reference can be done to
phenomenological model of Bari and Sivardi\`{e}re, who investigated the activation from the LS
ground state to excited HS states including the magnetoelastic coupling between the two species
\cite{RefBari1972PRB5_4466}. Depending on the model parameters, they obtained a succession of
three phases in coherence with experimental findings - the mixed LS/HS states with gradually
increasing HS weight at low temperatures, the disproportionation of HS-poor and HS-rich sites in
the intermediate range, and a re-entered homogeneous phase of LS/HS admixed states at high
temperatures. Very recently this scenario of spin-state transitions in \laco\ based on two states
only was supported in a study of a two-orbital Hubbard model with crystal-field splitting using
DMFT calculation by Kune\v{s} and K\v{r}\'{a}pek \cite{RefKunes2011PRL106_256401}. They
demonstrated that the existence of disproportionated phase and subsequent homogenization combined
with closing of the charge gap can be understood considering the on-site interactions of valence
electrons and their itinerancy. This suggests that the physics of \laco\ is primarily of
electronic (fermionic) origin, and elastic interactions between different spin-state species have
just a stabilizing effect.

The present study is undertaken with an aim to resolve by experimental and theoretical means the
role of electronic and elastic interactions in the stabilization of various Co$^{3+}$ spin states
in the \laco\ type related systems, in particular in \laco-\larh\ solutions. The studies on
Rh-containing compounds were not frequent in the past, mainly due to high cost of rhodium, which
limits potential applications. Nevertheless, since Rh$^{3+}$ of $4d^6$ configuration is
isoelectronic with Co$^{3+}$ as regards the valence electrons, and is exclusively in low-spin
state, the substitution is helpful in the fundamental research of cobalt spin states and
experienced an increased interest in the last years. Complete \lacorh\ solid solution was recently
studied in Ref.~\cite{RefLi2010JSSC183_1388} with focus on transport properties at high
temperature and in Ref.~\cite{RefAsai2011JPSJ80_104705} with focus on magnetic properties.
Influence of small Rh doping on the spin state transition in \laco\ was presented in
\cite{RefKyomen2003PRB67_144424}.

Our work revisits this system, focusing to compositions with small doping of Rh ($x=0-0.08$),
including for completeness also the $x=0.2$ and 0.5 samples. The experiments are complemented with
GGA+U electronic structure calculations using the supercell with regular orderings of Co and Rh
ions corresponding to doping $x=1/16=0.0625$, 1/2 and 15/16=0.9375. The method applied allows us
to investigate not only various valence and spin states in the model structures but also to
optimize local bond lengths and angles that stabilize the given configuration. The role of
temperature is simulated by a change of unit cell volume. In agreement with the experimental
findings, the calculations support the significant HS populations preserved in the \lacorh\
systems down to the lowest temperature, and suggest that the stabilization of HS Co$^{3+}$ states
compared to undoped \laco\ is to much extent due to elastic interactions. Namely, the low-energy
configurations are based on a nearest neighbor correlations of LS Co$^{3+}$ states of smaller size
and HS Co$^{3+}$ or LS Rh$^{3+}$ states of larger size in a kind of the double perovskite \abbo\
structure.

\section{Experimental}

Polycrystalline samples \lacorh\ with $x=0$, 0.02, 0.04, 0.08, 0.2 and 0.5, were prepared
according to the following procedure. Hot water solution of metal ions prepared by
decomposition/dissolution of La$_{2}$O$_{3}$, RhCl$_{3}$.H$_{2}$O, and
Co(NO$_{3}$)$_{2}$.6H$_{2}$O in 30\% HNO$_{3}$ was mixed with in advance prepared citric acid (CA)
- diethylenglycol (EG) water solution. The molar ratio CA:EG was 1:1 and the molar ratio
(CA+EG):metal ions was also 1:1. Clear, transparent and voluminous aerogel was obtained after
evaporation of water at 140\stc. The gel was pulverised in a mortar, dried, pyrolyzed, and organic
carbon residue were removed by proper heat treatment at temperatures 180-450\stc\ in a chamber
furnace under static air atmosphere. After careful mixing and grinding, the final crystallization
of \lacorh\ perovskite phase has been done by annealing at 1200\stc\ for 100~hours in air
atmosphere.

X-ray powder diffraction using a Bruker D8 diffractometer (CuK$\alpha$ radiation, SOL-X energy
dispersive detector, $2\theta$ range 20\st~-~150\st) was employed to determine the phase
compositions and lattice and structural parameters. The diffraction patterns were analyzed with
the Rietveld method using the FULLPROF program \cite{RefFullProf_PhysicaB}.

The magnetic properties were measured by means of a SQUID magnetometer MPMS-7 (Quantum Design)
over the temperature range $4.2-300$~K and a magnetosusceptometer DSM~10 (Manics) at elevated
temperatures up to 800~K.

Thermal conductivity, thermoelectric power and electrical resistivity were measured using a
four-probe method with a parallelepiped sample cut from the sintered pellet. The electrical
current density varied depending on the sample resistivity between $10^{-1}$~A/cm$^2$ (metallic
state) and $10^{-7}$~A/cm$^2$ (insulating state). The measurements were done on sample cooling and
warming using a close-cycle cryostat working down to 3~K.

\section{Method of calculation}

The calculations were done with the WIEN2k program \cite{RefWien2k}. This program is based on the
density functional theory (DFT) and uses the full-potential linearized augmented plane wave
(FP~LAPW) method with the dual basis set. The core states were defined as an electronic
configuration (Kr, $4d^{10}$) for La, (Ar, $3d^{10}$) for Rh, (Ne, $3s^2$) for Co and as (He) for
O atoms. The valence states included $3p$, $3d$ and $4s$ orbitals for Co, and $4s$, $4p$, $4d$ and
$5s$ for Rh.

All calculations were spin-polarized. For the exchange correlation potential the GGA form was
adopted \cite{RefPerdew1996PRL77_3865}. The radii of the atomic spheres were taken 2.3~a.u. for
La, 2.05~a.u. for Rh, 1.95~a.u. for Co and 1.55~a.u. for O. To improve the description of Co~$3d$
and Rh~$4d$ electrons we used the GGA+U method, which corresponds to $LDA+U$ method described in
\cite{RefAnisimov1993PRB48_16929,RefLiechtenstein1995PRB52_5467} with the GGA correlation
potential instead of $LDA$. The values of Coulomb and exchange parameters were used the same as in
the previous studies, $U=2.7$~eV and $J=0$~eV, for both Co and Rh
\cite{RefKnizek2006JPCM18_3285,RefKnizek2009PRB79_014430}.
%Coulomb interaction $U=2.7$~eV and intraatomic exchange $J=0$~eV
The atom positions were optimized by minimization of the calculated forces on the nuclei
\cite{RefTran2008CPC179_784}.

The calculations were performed in order to test two specific cases - small doping of Rh in \laco\
(and similarly small doping of Co in \larh) and $x=0.5$. For the first case, an ordered
arrangement of 1~Rh and 15~Co atoms was constructed ($x=1/16$), simulating a rather isolated Rh
atom in Co matrix. The model is characterized by intersected Rh-Co-Co-Co-Rh-$\dots$ chains running
along three pseudocubic axes, and a body centered supercell $2\sqrt{2}a_p \times 2\sqrt{2}a_p
\times 4a_p$ is used in order to include the octahedral tilting ($a_p$ is a lattice parameter of
the cubic perovskite cell). An analogical model structure for half doping $x=1/2$ is of double
perovskite type with 1:1 rock-salt arrangement of Co and Rh ions in the $\sqrt{2}a_p \times
\sqrt{2}a_p \times 2a_p$ cell, characterized by intersected Rh-Co-Rh-$\dots$ chains. The inversion
symmetry operation was only retained in both cells.

\section{Results}

%\subsection{X-ray diffraction}

%================= FIGURE =========================================================
\begin{figure}
\includegraphics[width=0.90\columnwidth,viewport=5 250 580 820,clip]{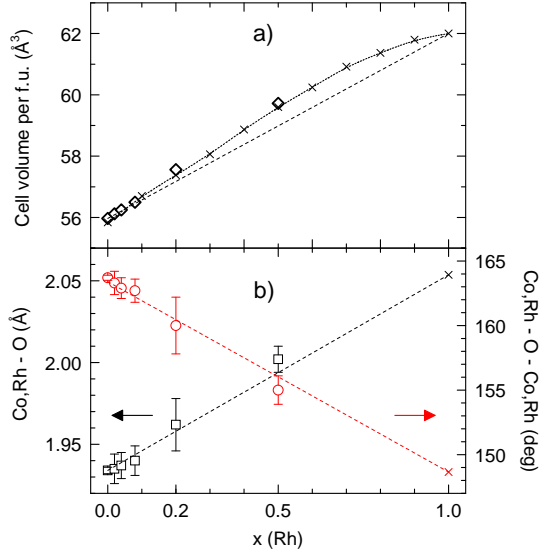}
\caption{(Color online) (a) Unit cell volume per formula unit (error bars are smaller than symbol
size) dependence on $x$ of \lacorh\ ($\diamond$).  The straight dashed line is a linear
interpolation between $x=0$ and 1. (b) Average Co,Rh-O bond lengths ($\square$) and Co,Rh-O-Co,Rh
angles (\textcolor{red}{$\circ$}) dependence on $x$. Corresponding values determined in
Ref.~\cite{RefLi2010JSSC183_1388} are shown for comparison ($\times$).} \label{figlatt}
\end{figure}
%==================================================================================

The symmetry of the unit cell is changed from rhombohedral \rtric\ for small doping $x=0-0.08$ to
orthorhombic $Pbnm$ for $x=0.2-1$ in agreement with previous studies
\cite{RefLi2010JSSC183_1388,RefAsai2011JPSJ80_104705}. The average bond distance Co,Rh-O increases
and average bond angle Co,Rh-O-Co,Rh decreases as more rhodium is inserted in the structure,
\textit{i.e.} the octahedral tilting is enhanced with increasing doping, see Fig.~\ref{figlatt}b.
These observations are in agreement with larger ionic radius of Rh$^{3+}$ (0.665~\AA\ in six-fold
coordination) compared to Co$^{3+}$. Actually, the ionic radius of Co$^{3+}$ significantly depends
on its spin state (0.545~\AA\ in LS and 0.610~\AA\ in HS state), nevertheless Rh$^{3+}$ is larger
than Co$^{3+}$ in any spin state.

The structural evolution of Co,RhO$_6$ octahedra is manifested in the increase of cell volume with
Rh doping, see Fig.~\ref{figlatt}a. The dependence shows a positive deviation from linearity in
agreement with previous studies \cite{RefLi2010JSSC183_1388,RefAsai2011JPSJ80_104705}. This
deviation could also be detected in the dependence of bond distances and angles, although it is
mostly hidden within error bars because the accuracy of these values is much lower, since X-ray
diffraction is very accurate in determination of lattice parameters, but is not as much sensitive
to oxygen positions.

%\subsection{Transport properties}

%================= FIGURE =========================================================
\begin{figure}
\includegraphics[width=0.90\columnwidth,viewport=5 0 580 820,clip]{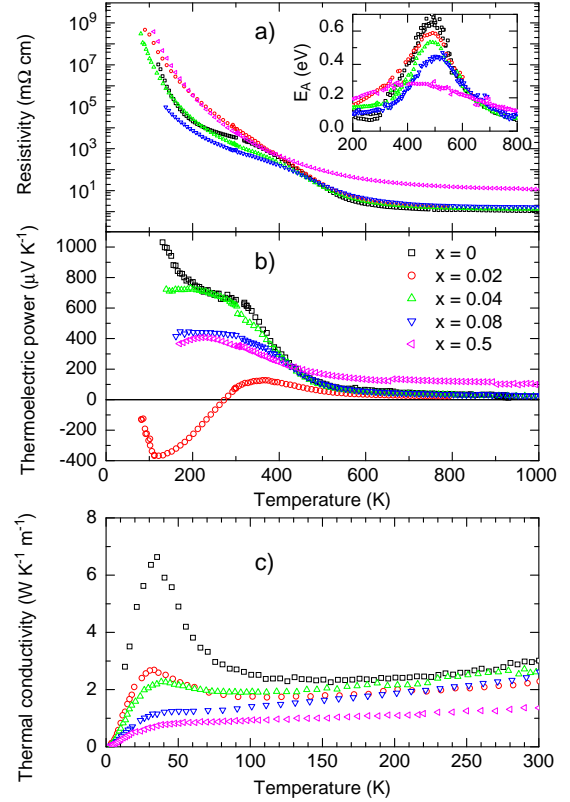}
\caption{(Color online) Resistivity, thermoelectric power and thermal conductivity dependence on
temperature of \lacorh.} \label{figsebres}
\end{figure}
%==================================================================================

Electric resistivity has semiconducting character at low temperature for all $x$, see
Fig.~\ref{figsebres}a. A transition to a more conducting state is observed above 400~K for small
Rh dopings, manifested by a peak in apparent activation energy around 500~K, see inset of
Fig.~\ref{figsebres}a. For $x=0.5$, only a broad hump is observed in the activation energy. High
thermoelectric power at low temperature indicates a low concentration of carriers, see
Fig.~\ref{figsebres}b. The absolute value of Seebeck coefficient ranges from 400 to 800~\muvk. A
decrease to a metallic-like value $\sim 30$~\muvk\ is observed above 400~K for small Rh doping. In
contrast, the Seebeck coefficient for $x=0.5$ saturates at a higher value of $\sim 100$~\muvk.
Thermal conductivity is decreasing and the phononic peak at low temperature is suppressed with
increasing Rh content, see Fig.~\ref{figsebres}c.

%\subsection{Magnetic properties}

%================= FIGURE =========================================================
\begin{figure}
\includegraphics[width=0.90\columnwidth,viewport=0 140 580 800,clip]{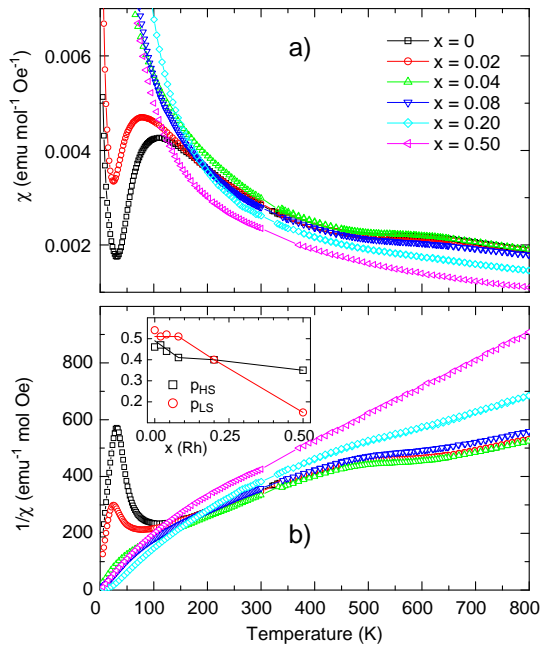}
\caption{(Color online) Molar magnetic susceptibility and inverse susceptibility of \lacorh. (The
spurious increase of susceptibility at low temperature for $x=0$ and 0.02 is a Curie term due to
minor impurity.) The inset shows calculated populations of HS and LS states.} \label{figsus}
\end{figure}
%==================================================================================

%================= TABLE ==========================================================
\begin{table}
\caption{The results of the magnetic susceptibility analysis in the middle temperature range
($150-300$~K): effective magnetic moment $\mu_{eff}$, Weiss $\theta$, and calculated ratio of the
Co cations in HS state ($p_{HS}$, $S=2$) and LS state ($p_{LS}$, $S=0$).}

\begin{tabular*}{\columnwidth}{@{\extracolsep{\fill}} l|rrrr}
\hline
$x$(Rh) &  $\mu_{eff}$ ($\mu_{B}$) & $\theta$(K) & $p_{HS}$ & $p_{LS}$ \\
\hline
0       & 3.31         & -203        & 0.46     & 0.54     \\
0.02    & 3.35         & -189        & 0.47     & 0.51     \\
0.04    & 3.25         & -140        & 0.44     & 0.52     \\
0.08    & 3.14         & -146        & 0.41     & 0.51     \\
0.2     & 3.10         & -159        & 0.40     & 0.40     \\
0.5     & 2.89         & -148        & 0.35     & 0.15     \\
\hline
\end{tabular*}

\label{tabsus}
\end{table}
%==================================================================================

Magnetic susceptibility measured in DC field of 10~kOe is displayed in Fig.~\ref{figsus}. The
low-temperature transition to LS state is shifted down in temperature by approx. 35~K for $x=0.02$
and it is suppressed for $x=0.04$. It means that inserting Rh into \laco\ destabilizes the purely
LS ground state of Co. As can be deduced from the slope of inverse magnetic susceptibility at the
lowest temperatures, a nonzero HS population is present starting from $x=0.04$. The poulation of
Co$^{3+}$ in HS state is estimated between $14-18$\% with maximum for $x=0.2$. The composition
$x=0.2$ deserves special attention, since the sample shows clear signatures of magnetic ordering
below 10~K, including the finite remanent magnetization in measurements of hysteresis loops, in
agreement with the recent paper of Asai~\etal\ \cite{RefAsai2011JPSJ80_104705}.

With increasing temperature, the slope of inverse susceptibility is changed as a result of the
excitation of further HS states and interactions among them. Finally, practically linear behavior
of inverse susceptibility is observed within the range $150-300$~K. At a higher temperature, the
samples $x=0 - 0.2$ exhibit a magnetic anomaly associated with the I-M transition. Its extent is
gradually diminished and shifted to slightly higher temperature with Rh doping, and for $x=0.5$
the transition completely vanishes, in agreement with the behavior of the anomaly in transport
data in the inset of Fig.~\ref{figsebres}a.

In order to characterize the room-temperature phases of \lacorh\ in a more quantitative manner,
let us discuss the results of the Curie-Weiss fit of inverse susceptibility within $150-300$~K,
which are summarized in Table~\ref{tabsus}. It is seen that the effective magnetic moment
$\mu_{eff}$ is enhanced for $x=0.02$ compared to $x=0$ and then it takes an opposite trend and is
decreasing with $x$. The number of Co$^{3+}$ in HS state ($p_{HS}$, $S=2$) can be calculated from
the $\mu_{eff}$ presuming Rh$^{3+}$ and remaining Co$^{3+}$ ions in the non-magnetic LS state. The
calculated $p_{HS}$ is reduced from 0.47 for $x=0.02$ to 0.41 for 0.08 while the number of
Co$^{3+}$ in LS state $p_{LS}$ is practically constant around 0.51, as though Rh were
predominantly replacing Co$^{3+}$ in HS state for the small doping range of $x$. On the other
hand, the extrapolation to $x=0$ gives $p_{HS}=0.49$, which is a somewhat higher value than
$p_{HS}=0.46$ actually determined for $x=0$. The origin of this discontinuity can be rationalized
as follows:

In the frame of the LS/HS thermal excitation model
\cite{RefKyomen2005PRB71_024418,RefKnizek2009PRB79_014430,RefKnizek2010JMMM322_1221}, the
population of HS states in pure \laco\ is limited to 50\%, since probability of nearest HS
neighbors is strongly suppressed. Because the excitations to HS state are dynamic and only a
short-range correlated, the actual limit should depend on the size of the correlated clusters of
alternating HS-LS states (or rather on the size of disordered boundaries between the clusters with
Co ions in LS states), and should be somewhat lower than 50\%. Therefore $\mu_{eff}$ corresponding
to $p_{HS} \sim 0.46$ is observed. We suppose, that the clusters of alternating HS-LS states,
which are stabilized by dilute Rh dopants representing immobile elastic defects, are larger than
the thermally induced dynamic clusters, hence the sum of Rh$^{3+}$ and HS Co$^{3+}$ ($x+p_{HS}$)
may approach closer the limit of 50\%.

The decrease of $\mu_{eff}$ with higher doping of Rh ($x>0.08$) is not so steep. In particular for
$x=0.5$, the calculated $p_{HS}$ only decreases to 0.35, while $p_{LS}$ is reduced down to 0.15.
We suppose, that the reason is in cumulation of large Rh ions, which creates a positive lattice
strain and thus supports HS state on nearby Co sites. This effect of Rh dopants for stabilization
of HS state at neighboring Co ions prevails for higher $x \rightarrow 1$, whereas the previously
mentioned effect of Rh ions replacing HS Co ions in LS/HS ordered regions and supporting HS state
on the next neighbor Co sites dominates for lower $x \rightarrow 0$. Nevertheless, in both cases
the HS states stabilized by Rh, unlike those obtained by thermal excitation, are preserved down to
the lowest temperature.

The dependence of cell volume on $x$ would be linear if the room temperature ratio $p_{HS}:p_{LS}$
remained constant at the original $x=0$ value for all $x$. Since this ratio becomes higher in the
doping region $x=0.2-0.5$ (and presumably also for $x>0.5$), the observed cell volume show a
marked positive deviations, see Fig.~\ref{figlatt}a.

The Weiss $\theta$ deduced from the susceptibility data over the range $150-300$~K are negative
for all $x$ and the absolute values are approximately decreasing with $p_{HS}$, as expected
regarding the antiferromagnetic interaction between Co$^{3+}$ in HS state.

%\subsection{Electronic structure calculation}

%================= FIGURE =========================================================
\begin{figure}[t]
\includegraphics[width=0.90\columnwidth,viewport=40 570 580 770,clip]{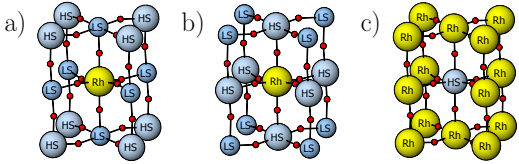}
\caption{(Color online) Examples of Co spin states configurations considered in the GGA+U
calculations.} \label{figRhLSHS}
\end{figure}
%==================================================================================

The previous GGA+U calculations evidenced, that two magnetic phases may exist in \laco\
perovskite, in addition to the non-magnetic LS ground state
\cite{RefZhuang1998PRB57_10705,RefKnizek2006JPCM18_3285,RefKnizek2009PRB79_014430}. The first one
is a result of gradual population of HS states conditioned by presence of the LS states at the
nearest Co sites. Nevertheless, some Co(HS) pairs do exist and are responsible for prevailing
antiferromagentic interactions. The HS state in LS matrix is further stabilized if Co(HS) site is
expanded (breathing-type) at the expense of neighboring Co(LS) sites. The mixed LS/HS phase tends
to a short-range ordered 1:1 arrangement, whose long-range formation is presumably prevented by
the entropy factor. The second phase consists of large IS clusters, which may exist within the
LS/HS phase and finally tend to a formation of uniform itinerant IS phase with ferromagnetic
coupling. The cell volume expansion destabilize the LS ground state and for certain critical
volumes the energy of LS/HS phase or IS phase become lower.

In the present GGA+U calculation, several spin states configurations of Co were tested using the
La$_{16}$Co$_{15}$RhO$_{48}$ supercell ($x=0.0625$), namely:
\begin{enumerate}
 \item All Co ions in LS state.
 \item All Co ions in IS state.
 \item Co in LS and HS state in 1:1 ratio, with the 6 nearest Co around Rh in LS state,
  see Fig.~\ref{figRhLSHS}a displaying the nearest neighbors (Co LS) and the next nearest neighbors (Co HS).
 \item Co in LS and HS state in 1:1 ratio, with the 6 nearest Co around Rh in HS state, see Fig.~\ref{figRhLSHS}b.
\end{enumerate}
Spin moment of Rh was allowed to vary, nevertheless it always converged to LS value. Cell volumes
were adopted to values expected for various doping by interpolation between \laco\ and \larh. Atom
positions determined at room temperature were used as the initial values and all refinable
coordinates were optimized. The cell volume was also decreased by 1-2\% to simulate the cell
contraction with temperature ($T\rightarrow 0$~K).

%================= FIGURE =========================================================
\begin{figure}
\includegraphics[width=0.90\columnwidth,viewport=0 470 580 820,clip]{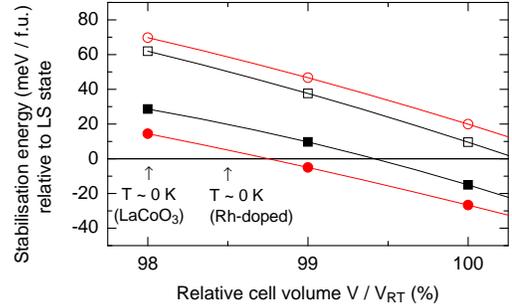}
\caption{(Color online) Stabilization energy of LS/HS (full symbols) and IS (open symbols)
configurations relative to LS configuration of \lacorh\ for $x=0$ ($\square$) and $x=1/16$
(\textcolor{red}{$\circ$}) in dependence on cell volume by GGA+U calculation.}
\label{FigLAPW_Volume}
\end{figure}
%==================================================================================

Fig.~\ref{FigLAPW_Volume} shows a comparison of the calculated energies for the LS/HS
configuration and homogenous IS phase relative to the LS phase, depending on the perovskite cell
volumes of \laco\ and LaCo$_{15/16}$Rh$_{1/16}$O$_{3}$.

The cell volume of \laco\ at 5~K is about 2\% smaller compared to room temperature
\cite{RefRadaelli2002PRB66_094408}, however only 1\% is related to temperature induced contraction
and another 1\% is related with the reduction of the Co$^{3+}$ size at the spin transition
\cite{RefKnizek2005EPJB47_213}. Since the spin transition in \lacorh\ is incomplete, the maximum
volume contraction is expected between $1-2$\% in this case. The relative cell volumes
approximately corresponding to $T=0$~K are indicated in the figure. In both cases, the lowest
energy state at room temperature (volume $V_{RT}$) is LS/HS, but the stabilization energy,
relative to LS state, is higher for Rh-doped compound. With the volume contraction simulating the
temperature decrease, the \laco\ system undergoes a crossover at about 99.4\% volume to the LS
ground state, while this crossover for LaCo$_{15/16}$Rh$_{1/16}$O$_{3}$ is around 98.7\% volume,
which is almost $T \rightarrow 0$~K in this case.

These calculations thus evidence that the LS/HS state is stabilized upon rhodium doping. The
driving force of this stabilization is the elastic energy. The ionic radius of Co$^{3+}$ strongly
depends on the spin state. This is confirmed by GGA+U structure optimization, which gives Co-O
bond lengths 1.92~\AA\ and 1.97~\AA\ for LS and HS states, respectively. The optimized Rh-O bond
length is around 2.02~\AA. Inserting Rh to Co matrix therefore brings about an increase of elastic
energy, which is relaxed by keeping the nearest Co neighbors in LS state and exciting the second
nearest Co neighbors do HS state, creating locally ordered cluster of large cations (Rh$^{3+}$ and
HS Co$^{3+}$) and small cations (LS Co$^{3+}$), see Fig.~\ref{figRhLSHS}a. The alternative LS+HS
state configuration, see Fig.~\ref{figRhLSHS}b, is evidently less favorable regarding the elastic
energy, as it was confirmed by the calculations.

Calculation for the $x=0.5$ within the La$_{4}$Co$_{2}$Rh$_{2}$O$_{12}$ supercell suggests that
the LS Co$^{3+}$ state is stable for $T \rightarrow 0$, energies of HS and LS state become
comparable at around the room temperature, and HS state is stabilized at elevated temperatures,
\textit{i.e.} upon further increase of the unit cell volume.

The calculation simulating the rhodium-rich limit $x \rightarrow 1$, \textit{i.e.} using the
La$_{16}$CoRh$_{15}$O$_{48}$ supercell (see Fig.~\ref{figRhLSHS}c), shows that in this case the HS
state of Co is more stable than LS state. This result can be understood considering that isolated
LS Co$^{3+}$ in the lattice of much larger Rh$^{3+}$ ions would represent an elastic defect of
large energy cost.

The results of GGA+U calculations are in agreement with magnetic susceptibility analysis. For
small doping the Rh atoms stabilize the HS states of cobalt at the next-nearest sites due to the
elastic energy, therefore the low-temperature transition to pure LS state is suppressed. But at
the same time, Rh ions actually occupy places which would be otherwise available for HS Co$^{3+}$,
thus the saturated number of Co in HS state is lowered with the doping and number of Co in LS
state is retained.

According to LS-LS/HS-IS model, the high temperature spin transition is based on formation of
uniform IS phase. Thus the destabilization of the IS state by Rh doping
(Fig.~\ref{FigLAPW_Volume}) is in agreement with suppression of this transition evidenced by
magnetic data.

%================= FIGURE =========================================================
\begin{figure}
\includegraphics[width=0.90\columnwidth,viewport=0 470 580 820,clip]{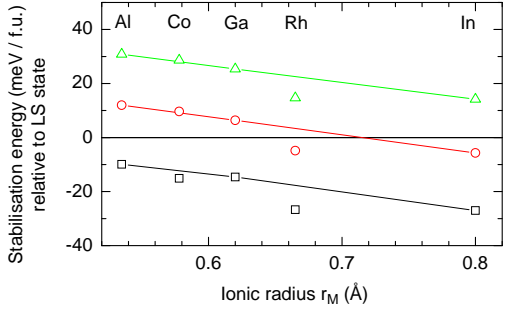}
\caption{(Color online) Stabilization energy of LS/HS relative to LS configuration of \lacotm\
(M~=~Co, Rh, Al, Ga and In) for $x=1/16$ in dependence on ionic radius $r_M$ (the average for
Co$^{3+}$ in LS and HS state is used for $r_{Co}$), by GGA+U calculation for room-temperature cell
volume $V_{RT} $($\square$), $V/V_{RT}=99\%$ (\textcolor{red}{$\circ$}) and $V/V_{RT}=98\%$
(\textcolor{green}{$\triangle$}).} \label{FigLAPW_rB}
\end{figure}
%==================================================================================

In order to investigate the relative role of elastic energy associated with larger Rh$^{3+}$ size
and purely electronic effects, additional simulations have been performed for $x=1/16$ doping of
other isovalent dopants with different ionic radii - Al$^{3+}$ ($r_{Al}=0.535$), Ga$^{3+}$
($r_{Ga}=0.62$) and In$^{3+}$ ($r_{In}=0.80$). Cell volumes were adopted to expected values for
$x=1/16$ doping by interpolation between \laco\ and respective LaMO$_3$. Atom positions determined
at room temperature were used as the initial values and all refinable coordinates were optimized.
The cell volume was also decreased by 1-2\% to simulate the cell contraction with temperature. The
results are presented in Fig.~\ref{FigLAPW_rB}. It is seen that the stabilization energy of the
LS/HS configuration with respect to pure LS phase is proportional to the ionic radius of the
doping cation in the Al, Ga and In doped systems, but in the case of Rh there is an additional
energy gain, obviously due to electrons of unfilled shell $4d$ and their covalency.

%================= FIGURE =========================================================
\begin{figure}
\includegraphics[width=0.90\columnwidth,viewport=5 80 580 820,clip]{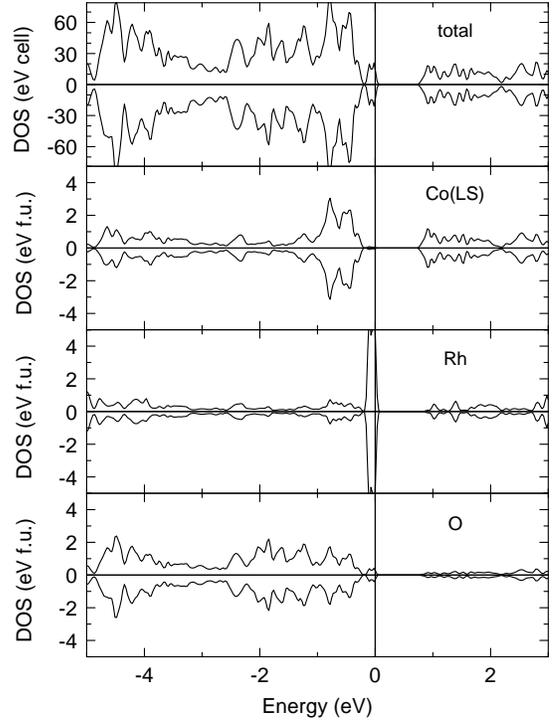}
\caption{Density of states for LS Co configuration of LaCo$_{15/16}$Rh$_{1/16}$O$_{3}$.}
\label{FigDOSLL}
\end{figure}
%==================================================================================

%================= FIGURE =========================================================
\begin{figure}
\includegraphics[width=0.90\columnwidth,viewport=5 80 580 820,clip]{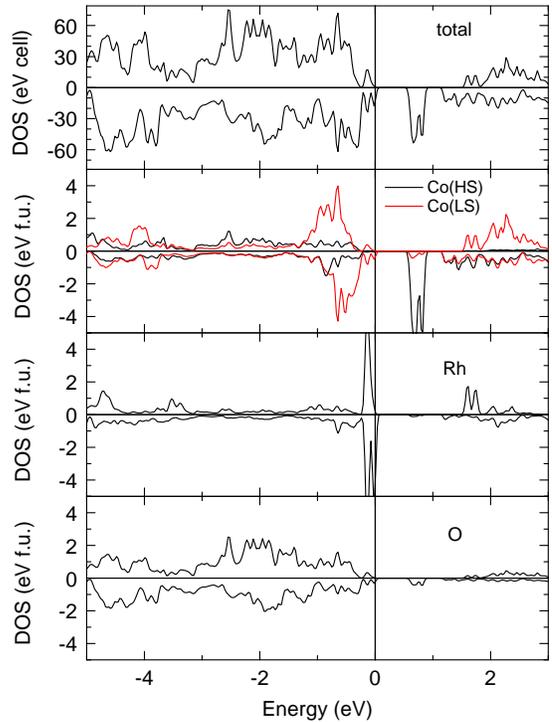}
\caption{(Color online) Density of states for LS/HS Co configuration of
LaCo$_{15/16}$Rh$_{1/16}$O$_{3}$.} \label{FigDOSHL}
\end{figure}
%==================================================================================

Total and atom projected density of states (DOS) of \lacorh\ ($x=1/16$) for LS spin configuration
is displayed in Fig.~\ref{FigDOSLL}. The character of DOS is insulating with a gap about 0.8~eV.
The narrow states $0-0.25$~eV below $E_F$ correspond to Rh($t_{2g}$), whereas Co($t_{2g}$) and
O($p$) states are situated lower in energy between $0.25-1$~eV. The states just above Fermi level
are mainly of Co character. DOS of LS/HS spin configuration is displayed in Fig.~\ref{FigDOSHL}.
The insulating character of DOS is retained with a slightly decreased gap about 0.5~eV. The narrow
states corresponding to Rh($t_{2g}$) are also within the energy range $0-0.25$~eV below $E_F$, and
the states of Co-LS($t_{2g}$) and O($p$) could be found lower in energy between $0.25-1$~eV. A
sharp peak above Fermi level belongs to Co-HS($t_{2g}$) band.

%\subsection{Electronic structure calculation - AIM}

It was assumed in the previous discussion, that there is no charge transfer between Co$^{3+}$ and
Rh$^{3+}$. To estimate quantitatively charge equilibria in our GGA+U calculations we use Atoms In
Molecule (AIM) concept of Bader \cite{RefBaderAIM}. In this approach the unit cell is divided into
regions by surfaces that run through the minima in the charge density. The charge on a given site
is obtained by integrating the electronic density within these regions. The advantage of this
method is that the analysis is based solely on the charge density, so it is independent on the
basis set and atomic spheres used.

The ionic charges calculated by the AIM method differ from the formal valencies due to
hybridization between cations and oxygen. The ratio of the AIM and ideal charge may be regarded as
the degree of hybridization. Therefore, a different AIM charge of cations with nominally equal
valencies can be solely caused by a different degree of hybridization with oxygen, and cannot be
directly considered as a charge transfer between the cations. Instead, we compare calculated AIM
charge of Co in undoped and doped structures. The AIM charges of cobalt in \laco\ are within
$1.40-1.64$ depending on the its actual spin state, since the hybridization with oxygen is
predominantly based on $e_g$ orbitals and thus depends strongly on their occupation, which is
different for each spin state. Essentially the same AIM charges of Co are obtained for doped
structure \lacotm\ ($x=1/16$), in spite of the various values calculated for the substituting
cations M, namely Rh (1.33), Al (1.98), Ga (1.94) and In (1.88). It means that the difference
between AIM charge of nominally Co$^{3+}$ and Rh$^{3+}$ in \lacorh\ solely account for the
different degree of hybridization of Rh and Co, but not present any ground for the charge transfer
between Co and Rh.

\section{Conclusion}

The present experiments supported by GGA+U electronic structure calculations provide strong
arguments that the transition metal ions in the \lacorh\ solid solutions remain in trivalent
states. The GGA+U calculations suggest that the elastic coupling of the nearest and next-nearest
cobalt neighbors of the inserted Rh dopant is important in stabilization of the spin-state
configurations. The Rh$^{3+}$ ions are always in the non-magnetic LS configuration, while the
Co$^{3+}$ ions may vary, depending on composition and temperature, between the non-magnetic LS and
paramagnetic HS local configurations. In distinction to pure \laco, in which HS Co$^{3+}$ states
only appear at increased temperature through a thermal activation process, the Rh doped systems
starting from $x=0.04$ exhibit certain number of stable HS Co$^{3+}$ species already in the ground
state (up to 18\% for $x=0.2$). The HS population increases gradually with increasing temperature
and reaches a saturation above 150~K, similarly to what is observed in undoped \laco. The
resulting phases, based on the LS/HS disproportionated cobalt sites, persist at least up to room
temperature. A gradual transformation to a uniform state with metallic-like conductivity is
evidenced at elevated temperatures for samples with $x<0.5$.

The main characteristics of the room-temperature phase of \laco\ is the LS/HS population close to
the 1:1 ratio. This can be understood as a result of cooperative effects, which prefer a regular
ordering of smaller LS Co$^{3+}$ ions and larger HS Co$^{3+}$. Nevertheless, only short-range
LS/HS correlations are anticipated considering the entropy reasons. Similar ordering tendencies
also exist in the doped systems but the mechanism is different. The primary HS population arises
from the elastic strain associated with presence of immobile Rh dopants. The strain is released by
stabilization of LS states on the nearest cobalt neighbors and HS states on the next nearest ones.
These clusters grow as additional HS states are thermally activated and tend to a saturation in
which small LS Co$^{3+}$ ions alternate with HS Co$^{3+}$ and LS Rh$^{3+}$ ions of larger size.
The stability of such arrangement can be related to two factors. The first one is the known
property of double perovskite structure \abbo\ to adapt to very different radii of $B$ and $B'$
cations. This is a kind of elastic energy optimization, which is effective not only for the
rhodium but also for other large isovalent dopants like In. The second mechanism is of purely
electronic origin and is associated with incomplete character of the $3d$ and $4d$ shells of Co
and Rh ions.

\textbf{Acknowledgments}. This work was supported by Project No.~202/09/0421 of the Grant Agency
of the Czech Republic.

%-KTeXBib[AUTHOR, \\ CRLF TITLE, \\ CRLF JOURNAL \textbf{VOLUME}, PAGES (YEAR). KURL KFILE]
%-KTeXBib[AUTHOR, \\ CRLF TITLE, \\ CRLF JOURNAL \textbf{VOLUME}, PAGES (YEAR).]
%\bibliography{Clanek_LaCoRhO3}

\end{document}